\PassOptionsToPackage{prologue,dvipsnames,table}{xcolor}
\documentclass[sigconf,nonacm]{acmart}
\AtBeginDocument{%
  \providecommand\BibTeX{{%
    \normalfont B\kern-0.5em{\scshape i\kern-0.25em b}\kern-0.8em\TeX}}}

\copyrightyear{2024} 
\acmYear{2024} 
\setcopyright{acmlicensed}\acmConference[WWW '24 Companion]{Companion Proceedings of the ACM Web Conference 2024}{May 13--17, 2024}{Singapore, Singapore}
\acmBooktitle{Companion Proceedings of the ACM Web Conference 2024 (WWW '24 Companion), May 13--17, 2024, Singapore, Singapore}
\acmDOI{10.1145/3589335.3648305}
\acmISBN{979-8-4007-0172-6/24/05}

\usepackage{amsmath}
\usepackage{graphicx}

\usepackage{subfigure}
\usepackage{stfloats}
\usepackage[table]{xcolor}
\usepackage{makecell}
\usepackage{multicol}
\usepackage{multirow}
\begin{document}

\title{Modeling User Viewing Flow Using Large Language Models for Article Recommendation}

\author{Zhenghao Liu}
\affiliation{
  \institution{Northeastern University}
  \city{Shenyang}
  \country{China}}
\email{liuzhenghao@mail.neu.edu.cn}

\author{Zulong Chen}
\authornote{indicates equal contribution.}
\affiliation{
  \institution{Alibaba Group}
  \city{Hangzhou}
  \country{China}}
\email{zulong.czl@alibaba-inc.com}

\author{Moufeng Zhang}
\authornotemark[1]
\affiliation{
  \institution{Alibaba Group}
  \city{Hangzhou}
  \country{China}}
\email{moufeng.zmf@alibaba-inc.com}

\author{Shaoyang Duan}
\affiliation{
  \institution{Alibaba Group}
  \city{Hangzhou}
  \country{China}}
\email{shaoyang.duanshaoy@alibaba-inc.com}

\author{Hong Wen}
\affiliation{
    \institution{Unaffiliated}
    \city{Hangzhou}
    \country{China}}
\email{dreamonewh@gmail.com}

\author{Liangyue Li}
\affiliation{
    \institution{Alibaba Group}
    \city{Hangzhou}
    \country{China}}
\email{liliangyue.lly@alibaba-inc.com}

\author{Nan Li}
\affiliation{
    \institution{Alibaba Group}
    \city{Hangzhou}
    \country{China}}
\email{nan.li@taobao.com}

\author{Yu Gu}
\affiliation{
  \institution{Northeastern University}
  \city{Shenyang}
  \country{China}}
   \email{guyu@mail.neu.edu.cn}

\author{Ge Yu}
\affiliation{
  \institution{Northeastern University}
  \city{Shenyang}
  \country{China}}
    \email{yuge@mail.neu.edu.cn}

\renewcommand{\shortauthors}{Zhenghao Liu et al.}


\begin{abstract}

This paper proposes the u\textbf{S}er view\textbf{ING} f\textbf{L}ow mod\textbf{E}ling (SINGLE) method for the article recommendation task, which models the user constant preference and instant interest from user-clicked articles. Specifically, we first employ a \textit{user constant viewing flow modeling} method to summarize the user's general interest to recommend articles. In this case, we utilize Large Language Models (LLMs) to capture constant user preferences from previously clicked articles, such as skills and positions. Then we design the \textit{user instant viewing flow modeling} method to build interactions between user-clicked article history and candidate articles. It attentively reads the representations of user-clicked articles and aims to learn the user's different interest views to match the candidate article. Our experimental results on the Alibaba Technology Association (ATA) website show the advantage of SINGLE, achieving a 2.4\% improvement over previous baseline models in the online A/B test. Our further analyses illustrate that SINGLE has the ability to build a more tailored recommendation system by mimicking different article viewing behaviors of users and recommending more appropriate and diverse articles to match user interests. 

\end{abstract}
\maketitle

\section{Introduction}
Online platforms, such as Microsoft News, Zhihu, CSDN, and Blog, provide an opportunity for users to visit news information, exchange user ideas and learn professional knowledge~\cite{wu2023personalized,alchokr2021comparative}. However, the overload of information makes users exhaust their patience to find the real-needed information from massive articles. The article recommendation system is necessary for users to find related articles to satisfy user information needs~\cite{an2019neural,wang2020fine,Du2021POLARAO}. Lots of article recommendation models mainly focus on the news recommendation task~\cite{wu2020mind}, which learns the user interest from previously clicked news articles and returns related news articles for users.
\begin{figure}[t]
\centering
\includegraphics[width=0.85\linewidth]{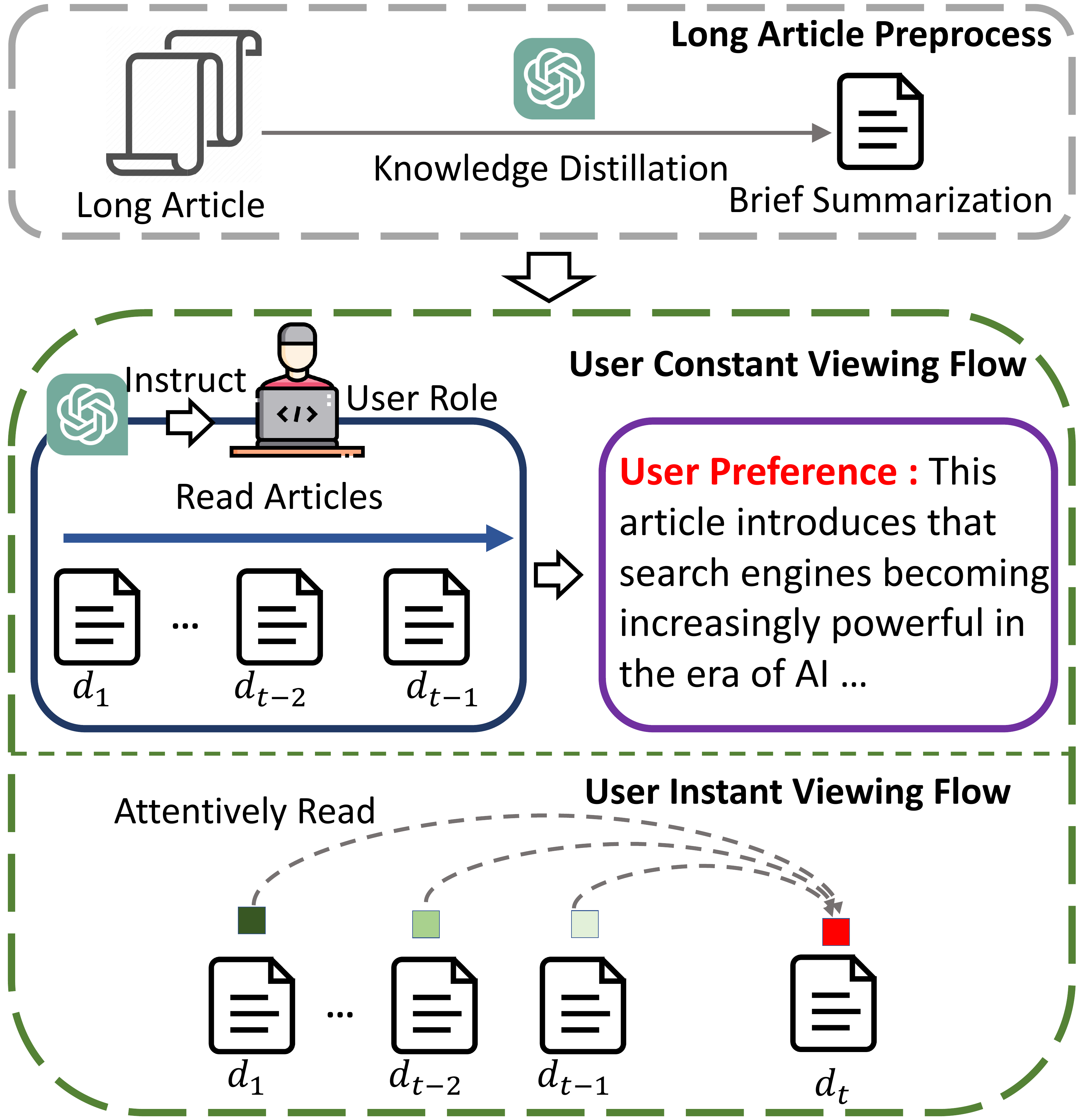}
\caption{The Architecture of User Viewing Flow Modeling (SINGLE) Method. It contains \textit{user constant viewing flow modeling} and \textit{user instant viewing flow modeling} methods.}
\label{fig:intro}
\end{figure}

Existing news recommendation models pay more attention to the personalized news recommendation techniques~\cite{qi2021personalized,bansal2015content,kompan2010content,liu2010personalized,zheng2018drn,wu2023personalized}. In the news recommendation task, a candidate news article may contain multiple aspects, topics and entities~\cite{liu2020kred,wu2019neural}, and the news recommendation models need to accurately match the candidate news with user interest~\cite{li2022miner,wu2019heterogeneous,wu2019neural,wang2020fine,wang2018dkn}. They learn the representations of news articles from their textual information, such as titles and bodies, and design the personalized attention mechanism to reweight different tokens for learning the news article representations~\cite{wu2019npa}. On the other hand, the sequence of user-clicked news articles usually contains some noise to represent the user's interest. Thus, lots of work considers the user personalized information to better capture user interest from the clicked articles, such as user IDs~\cite{wu2019npa,an2019neural}, user context information~\cite{de2018chameleon,zhang2019dynamic} and the user feedback~\cite{shi2021wg4rec,wu2019heterogeneous}. Large Language Models (LLMs), such as GPT3~\cite{openai2022chatgpt}, have shown strong abilities to summarize the user characteristics to better match items with user interests~\cite{lin2023can,wang2023zero,bao2023tallrec,zhang2023recommendation}.

In this paper, we propose a u\textbf{S}er view\textbf{ING} f\textbf{L}ow mod\textbf{E}ling (SINGLE) method to recommend articles for users. As shown in Figure~\ref{fig:intro}, we model the constant and instant viewing flows to better represent the interest of users, which helps to extract the general and instant user interests for recommendation. Specifically, we model the \textit{constant viewing flow} by prompting the LLM as a user to extract the user preferences and characteristics from visited articles. Then SINGLE models the \textit{instant viewing flow} to learn different interest views of users. It encodes the title and body of an article using BERT~\cite{devlin2018bert}, establishes the interactions between the representations of user-visited articles and candidate articles and attentively reads the representations of visited articles to conduct the user instant interest representation. Besides, SINGLE uses the text information from the titles and bodies of articles to better represent the article semantics. In this case, we design a summarization prompt by incorporating the article title as the gist, employ LLMs to extract the keywords from article bodies and then generate a brief paragraph to represent the article body.

The experiment results on the news recommendation dataset, MIND~\cite{wu2020mind} and the technique article recommendation dataset, Alibaba Technology Association (ATA), show the effectiveness of SINGLE. Furthermore, we deploy the SINGLE model on the online website of Alibaba. Notably, SINGLE achieves 2.4\% improvements on the online A/B test, which provides strong evidence to confirm its advantages in article recommendation tasks. Our further analyses show that the SINGLE model has the ability to mimic different article viewing behaviors and build a more effective article recommendation system. The user constant viewing flow modeling method learns general user interests by conducting more similar user representations with the clicked articles. Then the user instant viewing flow captures different user interests from the visited articles by assigning different weights to these user-clicked articles according to the candidate article. It can capture the consistent topic or topic shift from the clicked article sequence.

\section{Related Work}
The article recommendation task is a crucial task to improve the user experience on online reading platforms by reducing the information overload, which has drawn lots of attention from researchers and industry~\cite{das2007google,wu2023personalized,lin2014personalized,meng2023survey}. Existing article recommendation models mainly focus on the news recommendation task~\cite{wu2020mind}. They aim to learn the representations of news articles, model the characteristics and behaviors of users from their clicked article sequence, and then recommend news articles to satisfy the news information needs of users~\cite{lian2018towards,wang2011collaborative,wu2019neural,wu2019heterogeneous,wu2021empowering}.

Previous news recommendation models usually focus on better extracting features from the user-clicked article sequence. These models usually group the features of articles into different categories, represent different categories using randomly initialized embeddings and then try to model the relation between users and items by modeling interactions among different features. The Factorization Machine (FM) model~\cite{rendle2010factorization} uses pairwise inner products to better model the interactions among features. Wide\&Deep~\cite{cheng2016wide} and DeepFM~\cite{guo2017deepfm} further extract shallow and deep features by combining the advantages from both linear models and deep neural networks. DCN~\cite{wang2017deep} models feature interactions via deep and cross networks and builds high-degree interactions among features.

Recent news recommendation models regard the news recommendation task as a sequential recommendation task~\cite{yuan2023go,Geng0FGZ22,liu2023taste} and mainly focus on learning the representations of users and news articles and matching their representations for recommendation~\cite{wu2020neural,shi2021wg4rec,gao2018fine,zhang2019dynamic,okura2017embedding,kumar2017deep,kumar2017user}. These models learn the user representations by encoding the sequence of user-clicked news articles using different neural networks, such as RNN~\cite{hidasi2015session,liu2016context,donkers2017sequential}, CNN~\cite{tang2018personalized,an2019neural,Jiaxi2018Personalized,Fajie2018asc,yan2019cosrec}, and self-attention networks~\cite{kang2018self,sun2019bert4rec,wu2019nrms}. To better establish the interactions between users and news articles, many work learns news and user representations by capturing the personality characteristics using multi-view news article representation learning~\cite{wu2019neural}, such as using personalized attention networks to learn news representations according to user IDs~\cite{wu2019npa}, and utilizing multi-head self-attentions to encode news articles~\cite{wu2019nrms}. Some methods also enhance the news representations by incorporating external semantic information from knowledge graphs~\cite{wang2018dkn,liu2020kred,sheu2020context,lee2020news,liu2019news,qi2021personalized}. With the development of pretrained language models~\cite{devlin2018bert,liu2019roberta,raffel2020exploring}, lots of recommendation models~\cite{wu2021empowering,zhang2021unbert,Geng0FGZ22} thrive on the more effective text understanding ability of language models to learn the representations of news articles and users. Nevertheless, these models usually represent news articles with titles due to the max length limitation of pretrained language models~\cite{devlin2018bert,Vaswani2017attention,beltagy2020longformer,kitaev2020reformer}.

Large Language Models (LLMs)~\cite{touvron2023llama,openai2023gpt4,du2022glm} have shown strong emergent abilities~\cite{huang2022towards} and also benefit the recommendation systems~\cite{lin2023can}. \citet{wang2023zero} ask the GPT-3 model~\cite{openai2022chatgpt} to summarize the user characteristics from the user interaction history and then prompt the LLM to better recommend the item in the zero-shot setting. TALLRec~\cite{bao2023tallrec} fine-tunes the LLMs for recommendation by using the instruction tuning and recommendation tuning tasks. \citet{zhang2023recommendation} follow the self-instruct method~\cite{selfinstruct} to generate the instruct tuning dataset according to the user preference, intention and task forms. And then they leverage the self-instruct data to fine-tune the FlanT5 model~\cite{chung2022scaling} for recommendation. ChatREC~\cite{gao2023chat} and RecLLM~\cite{friedman2023leveraging} further employ the LLMs to enhance the conversational recommendation systems. They regard the recommendation systems as external tools, use LLMs to conduct the tool usage planning and generate more tailored responses.

 \section{Methodology}

\begin{figure*}[ht]
\centering
\includegraphics[width=\linewidth]{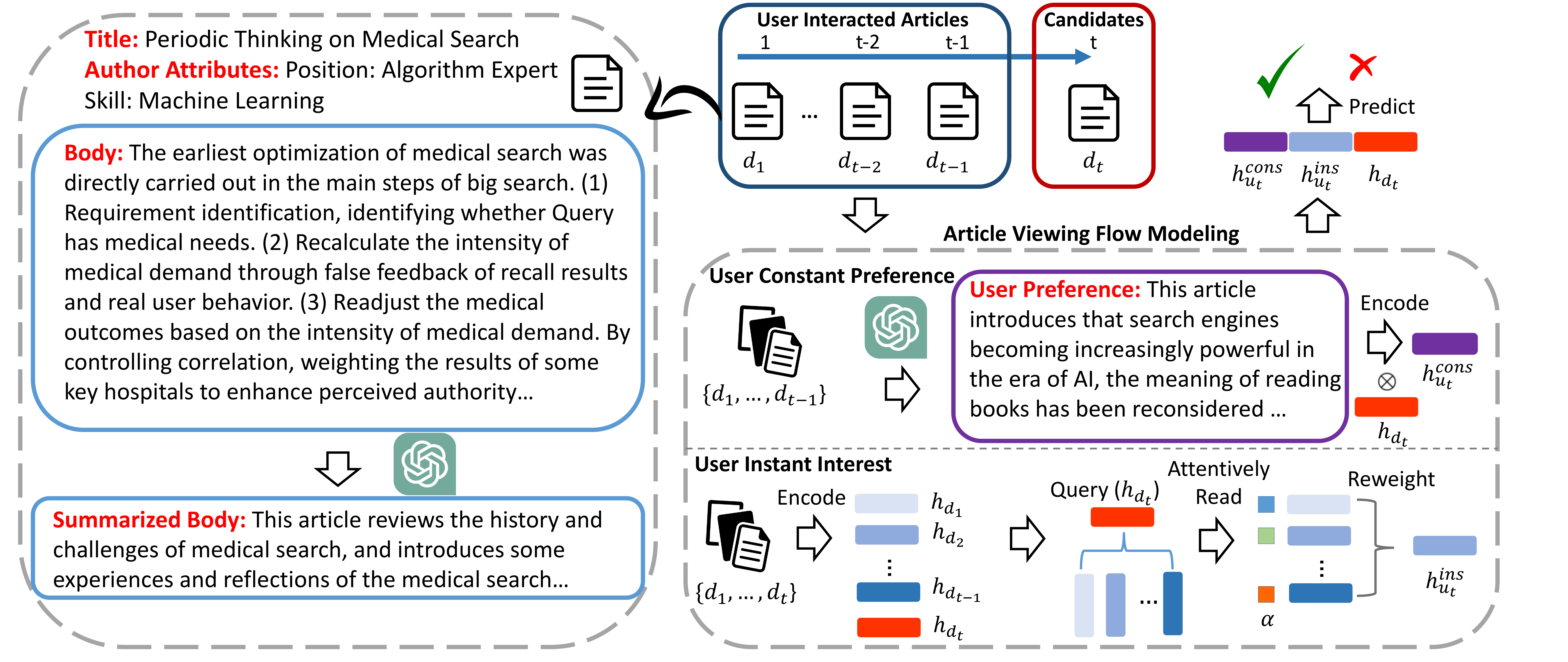}
\caption{The Framework of our User Viewing Flow Modeling (SINGLE) Method.}
\label{fig:model}
\end{figure*}
In this section, we introduce our User Viewing Flow Modeling (SINGLE) method. We first introduce the framework of SINGLE (Sec.~\ref{method:framework}). Then we describe our user constant and instant viewing flow modeling methods (Sec.~\ref{method:behavior}), which learn the general user characteristics and multi-view user interest. 

\subsection{Framework of SINGLE}\label{method:framework}
As shown in Figure~\ref{fig:model}, given the user viewed articles $D_{1:t-1} = \{d_1,..., d_{t-1}\}$, the article recommendation task aims to recommend the next article $d_t$ to satisfy the information need of users at the $t$-th time. In this subsection, we describe our article representation and encoding methods and then describe how to predict the next article according to the previously clicked documents $D_{1:t-1}$.

\subsubsection{Article Representation}
The article $d_i$ usually consists of the article title $d_i^t$, article body $d_i^b$ and article attributes $d_i^a$. Specifically, the article attributes contain the information of authors, such as department, gender and position, and the article view and click features. They provide additional ways to model the relevance between users and articles. The title and body of the article contain crucial semantics and help to understand the user behaviors via the text descriptions. However, existing article recommendation methods usually face the long text modeling problem, which derives from the max length limitation of existing language models. Thus, these models discard the article body and only use the article title to represent the article. It is evident that such an article representation method usually loses some critical semantics, which limits the effectiveness of these article recommendation systems.

To fully model the semantics of article $d_i$, we try to borrow the strong generalization ability of large language models (LLM) to reduce the text length of article bodies. Specifically, we design the article summarization instruction $instruct_s$ and employ LLMs to generate a brief summarization $d_i^{b*}$ for the article body $d_i^b$:
\begin{equation}\label{eq:summarization}
    d_i^{b*} = \text{LLM}(instruct_s, d_i^t, d_i^b),
\end{equation}
where $instruct_s$ is the summarization instruction. The length of summarized article body $d_i^{b*}$ is smaller than the length of raw article body $d_i^b$, \textit{e.g.} the average lengths of raw and summarized article bodies in the ATA dataset is 3,902 and 260, respectively.

Finally, we replace the raw article body $d_i^b$ with the summarized one $d_i^{b*}$ and the article $d_i$ can be represented with article title $d_i^t$, summarized article body $d_i^{b*}$ and article attributes $d_i^a$.

\subsubsection{Article Encoding} We can encode the document $d_i=\{d_i^a, d_i^t, d_i^{b*}\}$ as the article representation $h_{d_i}$:
\begin{equation}\label{eq:article_encode}
    h_{d_i} = h_{d_i}^a; h_{d_i}^t; h_{d_i}^{b*},
\end{equation}
where $;$ is the concatenation operation. $h_{d_i}^a$, $h_{d_i}^t$ and $h_{d_i}^{b*}$ are the embeddings of the attributes $d_i^a$, title $d_i^t$, and summarized body $d_i^{b*}$ of the $i$-th article $d_i$.

Specifically, we can get the title representation $ h_{d_i}^t$ and the body representation $h_{d_i}^{b*}$ by using BERT~\cite{devlin2018bert}:
\begin{equation}\label{eq:encode}
    h_{d_i}^t = \text{Linear}(\text{BERT}(d_i^t));
    h_{d_i}^{b*} = \text{Linear}(\text{BERT}(d_i^{b*})),
\end{equation}
where the BERT parameters are frozen during training. These article features $d_i^a$ are represented by the randomly initialized embeddings. Then we concatenate these attribute embeddings and employ a multi-layer neural network to get the final representation of article attributes $h_{d_i}^a$.

\subsubsection{Next Article Prediction}
Our SINGLE model learns the representations $h_{u_t}$ for the user by modeling user instant interest and constant preference from previously visited articles. Then we can predict the click rate of the candidate article $d_t$ at the $t$-th time:
\begin{equation}\label{eq:label_pred}
    P(y=1|D_{1:t-1}, d_t) =  \text{Sigmoid}(\text{Linear}(h_{u_t}; h_{d_t})),
\end{equation}
where $;$ is the concatenation operation and $h_{d_t}$ is the encoded representation of the article $d_t$. $P(y=0|D_{1:t-1}, d_t) = 1- P(y=1|D_{1:t-1}, d_t)$. Then we employ the cross entropy loss $\mathcal{L}$ to optimize the trainable parameters of our SINGLE model:
\begin{equation}
\mathcal{L} = \text{CrossEntropy} (y^*, P(y|D_{1:t-1}, d_t)),    
\end{equation}
where $y^*$ is the ground truth label of the $t$-th time predicted article. $y^*$ can be categorized into two types, including clicked ($y^*=1$) and not clicked ($y^*=0$). In the following subsection (Sec.~\ref{method:behavior}), we will further introduce the user behavior modeling method.

\subsection{User Viewing Flow Modeling}\label{method:behavior}
In this subsection, we model the user behaviors using the article clicked history $D_{1:t-1} = \{d_1,..., d_{t-1}\}$ and learn the user representation $h_{u_t}$. Specifically, we aim to mimic user instant interest and user constant preference by modeling the instant and constant viewing flows to recommend the article at the $t$-th time. We concatenate the user instant interest representations $h_{u_t}^{ins}$ and user constant interest representation $h_{u_t}^{cons}$ to represent users:
\begin{equation}\label{eq:gate}
    h_{u_t} = h_{u_t}^{ins}; h_{u_t}^{cons}.
\end{equation}
The user instant viewing flow modeling method and user constant viewing flow modeling method are introduced in Sec.~\ref{method:instant} and Sec.~\ref{method:constant}, respectively.

\subsubsection{User Instant Viewing Flow Modeling}\label{method:instant}
The users usually perform multi-view interests during viewing articles~\cite{wu2023personalized,wu2019npa} and the user instant interest indicates the rapid and immediate preference of users. 
During suffering the articles, some articles that belong to different topics can pique the curiosity of users and appeal to users to click these articles.

To model the instant interests of users, SINGLE models the instant viewing flow by regarding the potentially interacted article $d_t$ at the $t$-th time as a query and attentively reading the semantics from previously visited articles of the user $D_{1:t-1} = \{d_1,..., d_{t-1}\}$. Such an attention mechanism aims to extract some semantics from these visited articles $D_{1:t-1}$ according to the candidate document $d_t$, which can provide some supporting evidence to predict the next article.
Specifically, we first use Eq.~\ref{eq:article_encode} to get the article representations $H_{1:t-1} = \{h_1,..., h_{t-1}\}$ of visited documents $D_{1:t-1} = \{d_1,..., d_{t-1}\}$ and the representation $h_{d_t}$ of the potentially interacted document $d_t$ at the $t$-th time. Then the user instant interest representation $u^{ins}_t$ can be calculated:
\begin{equation}
h_{u_t}^{ins} = \sum_{i=1}^{t-1} \alpha_i \cdot h_{d_i},  
\end{equation}
where $\alpha_i$ is the attention score to weight the representations of visited document $d_i$:
\begin{equation}\label{eq:attention}
\alpha_i = \text{softmax}_i (h_{d_t} \cdot W \cdot (h_{d_i})^T),
\end{equation}
where $W$ is a learnable matrix. 

\subsubsection{User Constant Viewing Flow Modeling}\label{method:constant} 
Different from our user instant viewing flow modeling (Sec.~\ref{method:instant}), the user constant viewing flow aims to disentangle the general preference of users from different clicked articles and learn static user characteristics.

To model the user preference, we mimic the general user viewing behaviors using a token-level collaboration filter model. Specifically, we concatenate the text representations of previously clicked articles $D_{1:t-1} = \{d_1,..., d_{t-1}\}$, employ LLMs to capture some common keywords and capture the user characteristics from $D_{1:t-1}$:
\begin{equation}\label{eq:user_summarization}
    u^{cons}_t = \text{LLM}(instruct_u, D_{1:t-1}),
\end{equation}
where $instruct_u$ is the instruction to prompt LLMs to play a role in extracting the user characteristics from $D_{1:t-1}$. Then the constant user interest representation $h_{u_t}^{cons}$ can be encoded:
\begin{equation}
    h_{u_t}^{cons} = \text{BERT}(u^{cons}_t) \otimes h_{d_t},
\end{equation}
where $\otimes$ is the element-wise product. The BERT encoder shares the same parameters with the BERT encoder in the Eq.~\ref{eq:encode}.  We use the flow gate ($\otimes$) to control the semantics of the candidate article $h_{d_t}$ using the user constant interest.
\section{Experimental Methodology}
In this section, we introduce the datasets, evaluation metrics, baselines and implementation details of our experiments.
\begin{table}[t]
\begin{center}
\caption{\label{tab:data}Data Statistics of ATA and MIND.}
\begin{tabular}{l | r r | r r}
\hline 
\multirow{2}{*}{\textbf{Dataset}} &  \multicolumn{2}{c|}{\textbf{Data Info}} &  \multicolumn{2}{c}{\textbf{Split}}\\
&\textbf{\#User} & \textbf{\#Article} & \textbf{Train}  & \textbf{Test}  \\ \hline
ATA & 44,758 & 146,219 & 9,196,714  &  52,583 \\
MIND & 2,000 & 6,444 & 24,140 & 2,800 \\
\hline
\end{tabular}
\end{center}
\end{table}
\subsection{Datasets}
In our experiments, we use two datasets, ATA and MIND, to evaluate the effectiveness of different recommendation models. The data statistics are shown in Table~\ref{tab:data}.

\textbf{ATA.} The ATA dataset is collected from the Alibaba Technology Association (ATA), which was established in August 2012. It is an internal platform within Alibaba's technical ecosystem designed for engineers to exchange ideas and collaborate on technical matters, which helps to foster an engineer-focused culture and encourages technological innovation. ATA provides Alibaba's technical professionals with functions such as blogs, specialized topics, communities, and event organization.

\textbf{MIND.} MIND (Microsoft News Dataset)~\cite{wu2020mind} is a large-scale dataset open-sourced by Microsoft, which serves the research in news recommendation\footnote{\url{https://msnews.github.io/}}. The data collection originates from anonymous behavioral logs on the Microsoft News website. MIND consists of approximately 160,000 English news articles and over 15 million logs generated by one million users.

\subsection{Evaluation Metrics}
In our experiments, we follow previous work and evaluate the recommendation effectiveness of different models using AUC, MRR, nDCG@5 and nDCG@10 as our evaluation metrics.

We also utilize the UVCTR evaluation metric to evaluate the effectiveness of SINGLE, which is deployed on the ATA website.
\begin{equation}
\label{eq:uvctr}
\text{UVCTR}=\frac{\text{\#users who clicked articles}}{\text{\#users who visited the ATA homepage}}
\end{equation}

\subsection{Baselines}
We compare several baseline models with our SINGLE model. These models mainly focus on the news recommendation task.

\textbf{DCN~\cite{wang2017deep}.} DCN introduces a cross-network to better extract the item features using the deep neural network. The cross-network helps to extract the high-degree interaction across features to represent the items.

\textbf{DIN~\cite{zhou2018deep}.} The Deep Interest Network (DIN) establishes the interactions between candidate items and users to adaptively learn the representation of user interest from historically interacted items. It proposes a local activation unit to calculate the relevance between candidate items and previously interacted items. 

\textbf{NPA~\cite{wu2019npa}.} NPA represents news articles by encoding their titles. Then it uses the embedding of user ID to generate the query vector and further proposes a personalized attention network by using the word-level and article-level attention mechanisms.

\textbf{NAML~\cite{wu2019neural}.} The NAML model aims to learn multi-view representations for news articles. It encodes the titles and bodies of news articles using CNN. Then NAML learns the article representations by using a query embedding to attentively read the encoded token representations from the categories, titles and bodies of articles.

\textbf{NRMS~\cite{wu2019nrms}.} Instead of encoding news articles with CNN, NRMS further leverages the multi-head attention mechanism to encode the titles to represent news articles.

\textbf{FedRec~\cite{qi2020privacy}.} FedRec enhances classical federated learning recommendation systems to achieve privacy protection in news recommendations. FedRec uses a user model to learn user interests and leverages a model to learn news representations. Each user locally computes model gradients, encrypts them using Local Differential Privacy (LDP), and uploads them to the server.

\begin{table*}[t]
     \centering
     \caption{Overall Article Recommendation Performance of Different Models on ATA and MIND Datasets.\label{tab:overall}}
    \begin{tabular}{l|cccc|cccc}
    \hline
        \multirow{2}{*}{\textbf{Model}} & \multicolumn{4}{c|}{\textbf{ATA}} &\multicolumn{4}{c}{\textbf{MIND}}\\\cline{2-9}
        & AUC & MRR & nDCG@5 & nDCG@10 & AUC & MRR & nDCG@5 & nDCG@10 \\ \hline
        DCN~\cite{wang2017deep} & 67.69 & 40.81 & \textbf{50.48} & 61.01 & 63.43 & 40.40 & 47.56 & 57.67 \\ 
        DIN~\cite{zhou2018deep} & 67.19 & 39.90 & 49.30 & 60.30 & 63.01 & 40.94 & 47.56 & 58.44 \\ 
        NPA~\cite{wu2019npa} & 67.05 & 39.55 & 49.04 & 59.96 & 61.22 & 38.69 & 43.27 & 55.72 \\
         
        NAML~\cite{wu2019neural} & 66.85 & 39.22 & 48.60 & 59.61 & 64.30 & 41.12 & 48.06 & 58.12 \\         NRMS~\cite{wu2019nrms} & 67.28 & 40.11 & 49.45 & 60.47 & 64.75 & 41.69 & 48.16 & 58.80 \\ 
        FedRec~\cite{qi2020privacy} & 66.95 & 39.58 & 48.83 & 59.89 & 63.02 & 41.66 & 47.80 & 58.65 \\ 
    
        SINGLE & \textbf{68.06} & \textbf{41.01} & 50.40 & \textbf{61.28} & \textbf{66.50} & \textbf{44.22} & \textbf{52.26} & \textbf{60.97} \\ \hline
    \end{tabular}
\end{table*}

\subsection{Implementation Details}
This subsection describes the implementing details of SINGLE.

During training, we use batch normalization, set the dropout rate as 0.1 and set the batch size as 512. We employ Adam optimizer to optimize the model parameters and set the learning rate as 1e-5. Our model is trained about 600,000 steps and the early stop is used. The parameters of the BERT model are frozen and we initialize the BERT parameters using the checkpoints from the Chinese sentence embedding project\footnote{\url{https://www.modelscope.cn/models/damo/nlp_corom_sentence-embedding_chinese-tiny}}, which is trained using the DuReader dataset~\cite{Qiu2022DuReader}.

For the MIND dataset, our SINGLE model leverages gpt-3.5-turbo\footnote{\url{https://platform.openai.com/docs/models}} as the LLM module and uses different prompts to summarize information from the article bodies and extract the user preference from clicked articles. The article summarization instruction (Eq.~\ref{eq:summarization}) is ``This is an article about [category], please summarize it in a short sentence by piquing the reader's interest: [article\_body]''. Then SINGLE uses the instruction ``Please summarize the user's news browsing content. Here is the browsing history: [visited\_articles]'' to summarize the user constant preference.

For the ATA dataset, we employ ChatGLM-6B\footnote{\url{https://github.com/THUDM/ChatGLM-6B}} to implement the LLM module due to the privacy license. The article body summarization instruction ($instruct_s$) and user constant preference instruction ($instruct_u$) are ``Given an article, the title is [article\_title] and the article content is  [article\_body], please generate a 200-word summarization according to the article.'' and ``Given the visited articles: [visited\_articles],
I am a [position] from [organization], and my skills are [skill]. Please write a summary of about 150 words based on the visited articles.'', respectively.

\section{Evaluation Results}
In this section, we evaluate the effectiveness of our SINGLE model by showing its overall performance, conducting ablation studies, and evaluating the roles of different user visiting flow modeling methods. Finally, we conduct the online A/B test and case studies for further evaluation.

\subsection{Overall Performance}\label{sec:overall}
We first evaluate the recommendation effectiveness of SINGLE and show the model performance in Table~\ref{tab:overall}. Overall, our SINGLE model outperforms all baseline models, especially on the MIND dataset. It demonstrates the effectiveness of SINGLE in recommending more tailored articles to satisfy user needs in both news article and technique article recommendation tasks.

Among all baselines, the personalized news recommendation methods (NAML, NRMS and FedRec) usually present better performance than the feature-based article recommendation methods (DCN and DIN) on the MIND dataset. It demonstrates that such a personalized news recommendation modeling method is effective in capturing user interest for matching candidate news articles. On the contrary, these two groups of recommendation methods show almost the same performance on ATA dataset. The main reason may lie in that the ATA dataset contains different roles of users, such as algorithm expert, product manager and algorithm engineer, making it more difficult to model the universal interest representations for different kinds of users. SINGLE shows consistent improvements on both datasets, showing its advantages in modeling the interest of different users using our article viewing flow modeling method. 

\begin{table}
    \centering
    \caption{Ablation Studies. We evaluate the effectiveness of different modules of SINGLE on ATA dataset.\label{tab:ablation}}
    \resizebox{\linewidth}{!}{
    \begin{tabular}{l|cccc}
    \hline
        \textbf{Model} & \textbf{AUC} & \textbf{MRR} & \textbf{nDCG@5} & \textbf{nDCG@10} \\ \hline
        SINGLE & 68.06 & 41.01 & 50.40 & 61.28 \\\hline 
        w/o Summarized Bodies & 66.92 & 39.39 & 48.52 & 59.56 \\
        w/o Instant Flow & 67.19 & 40.11 & 49.01 & 60.45 \\
         w/o Constant Flow & 67.12 & 39.86 & 48.83 & 60.20 \\ 
        w/o Flow Gate  & 67.07 & 39.84 & 48.78 & 60.13 \\ 
        w/o $Instruct_u$ & 67.18 & 39.86 & 49.04 & 60.18 \\ \hline
    \end{tabular}}
\end{table}

\subsection{Ablation Studies}
In this experiment, we conduct ablation studies to show the roles of different modules in our SINGLE model.

As shown in Table~\ref{tab:ablation}, we first evaluate the effectiveness of LLMs in summarizing article bodies and illustrate the ability of LLMs in extracting knowledge from long texts. Then we present the effectiveness of different article flow modeling methods, including user constant viewing flow modeling and user instant viewing flow modeling. Finally, we conduct two models, SINGLE w/o Flow Gate and SINGLE w/o $Instruct_u$, to explore the roles of flow gating mechanism (Eq.~\ref{eq:gate}) and user profile promoting (Eq.~\ref{eq:user_summarization}) in extracting user constant preference from their clicked articles. 

The experimental results show that both user instant viewing flow and constant viewing flow modeling methods are effective in recommending the articles to users. It illustrates the potential article viewing behavior of users, which should match the general interest of users and also recommend some related and fresh articles to draw the attention of users. Compared with user instant viewing flow modeling~\cite{wu2019npa,wu2023personalized}, our user constant viewing flow modeling is more effective in recommending more tailored articles for users. It demonstrates that our constant viewing flow modeling is effective in filtering out noise from the article viewing history and better capturing user preferences and general interest.

In our SINGLE model, the effectiveness of our user constant viewing flow modeling method mainly derives from the strong ability of LLMs. During modeling the user constant viewing flow, the user role prompt $Instruct_u$ utilizes a template to capture the skill and position information of users. It shows the effectiveness in prompting LLMs to extract the characteristics of users from the user-clicked articles. On the other hand, our flow gate mechanism plays a critical role in combining the user constant representations for article recommendation. The flow gate mechanism uses an element-wise operation to learn a new view of the candidate article, which uses the user constant representation to control the information semantics of the candidate article to predict the user click probability.

\begin{figure}[t]
 \subfigure[\textbf{User A:} He/She is a front-end technology expert with the skill of front-end architecture and development.] { \label{fig:topic:a}
\includegraphics[width=\linewidth]{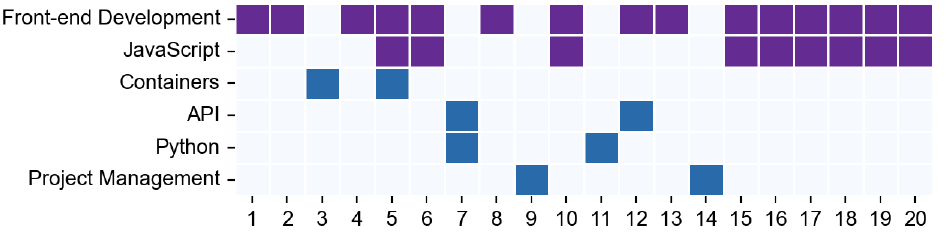}}
 \subfigure[\textbf{User B:} He/She is a software engineer with the skill of server-side architecture and development.] { \label{fig:topic:a}
\includegraphics[width=\linewidth]{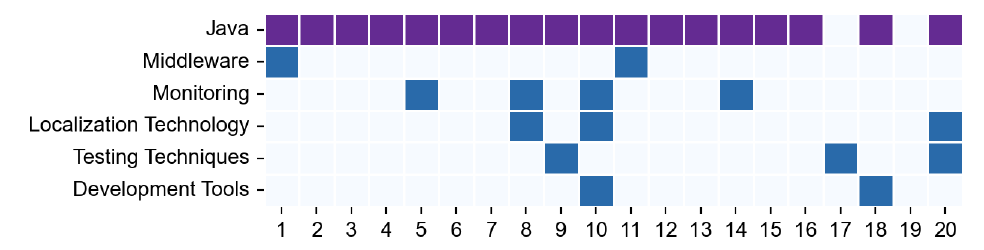}}
 \subfigure[\textbf{User C:} He/She is a technical expert with skills in operations and high availability.] { \label{fig:topic:a}
\includegraphics[width=\linewidth]{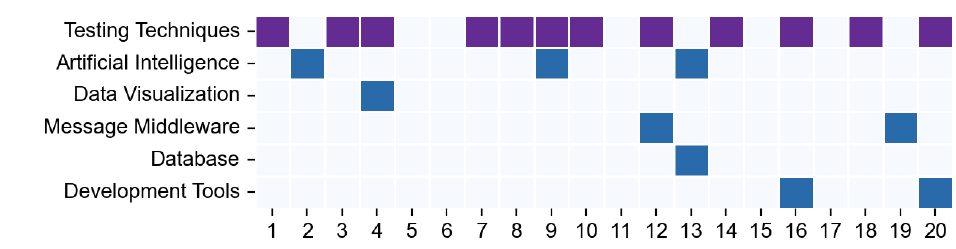}}
\caption{The Topic Distribution of User Clicked Articles. We randomly choose three different users from the ATA dataset to show the topic distribution of 20 clicked articles. The articles that are annotated with the purple and blue indicate the constant interest and instant interest of users, respectively. }
\label{fig:topic}
\end{figure}
\subsection{The Roles of Different User Viewing Flow Modeling Methods}
In this subsection, we evaluate the roles of different user viewing flow modeling methods in learning user behaviors. We first show the topic distribution of the clicked articles of different users. Then we explore the mechanism of our user instant and constant viewing flow modeling by showing the similarity scores between the representations of the user and the visited articles.

\begin{figure}[t]
\centering
\includegraphics[width=0.95\linewidth]{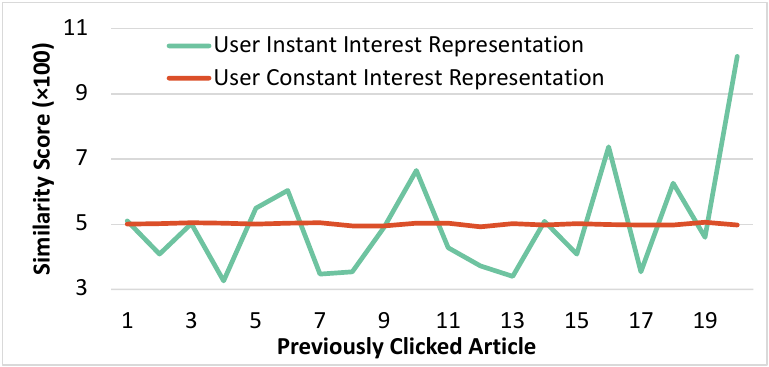}
\caption{The Similarity Scores between Representations of Users and Previously Clicked Articles. We calculate the average cosine similarity scores between user representations and clicked articles at different time steps. The user instant and constant representations are conducted by our instant and constant viewing flow modeling methods.}
\label{fig:weight}
\end{figure}
In Figure~\ref{fig:topic}, we choose three users to show the topic distributions of their clicked articles. Overall, these users usually prefer to click articles that are related to JavaScript, Java and Testing Techniques, showing the general interest of users. These article topics are associated with the skill and position of the users, which supports the motivation of our user constant interest modeling method. On the other hand, the user also clicked the articles that are on different topics, illustrating that the users usually have different interest points that can draw their attention~\cite{wu2019npa,wu2019neural}. Both article visiting behaviors are critical to building a tailored recommendation system. In this case, modeling user constant and instant visiting flows can improve user experience by recommending more appropriate and diverse articles. 

Then we show the similarity between the representations of users and clicked articles in Figure~\ref{fig:weight}. The user constant representations conduct consistent similar scores with the user clicked articles as different time steps, showing that our user constant viewing flow modeling method learns general semantics from these clicked articles. It helps to extract the static characteristics of users and recommend appropriate documents, which share a similar topic and match the general interest of users. On the contrary, the user instant interest representations show a distinct modeling behavior. The user instant interest representations are more similar to some of the article representations, especially these later user interacted articles, showing the users usually have different interest views that can match the articles of different topics. Such an instant interest modeling method makes the recommendation system return some diverse articles to satisfy the interest of users, which helps to avoid dull and more general recommendation results.

\subsection{Online A/B Test}

\begin{figure}[t]
\centering
\includegraphics[width=\linewidth]{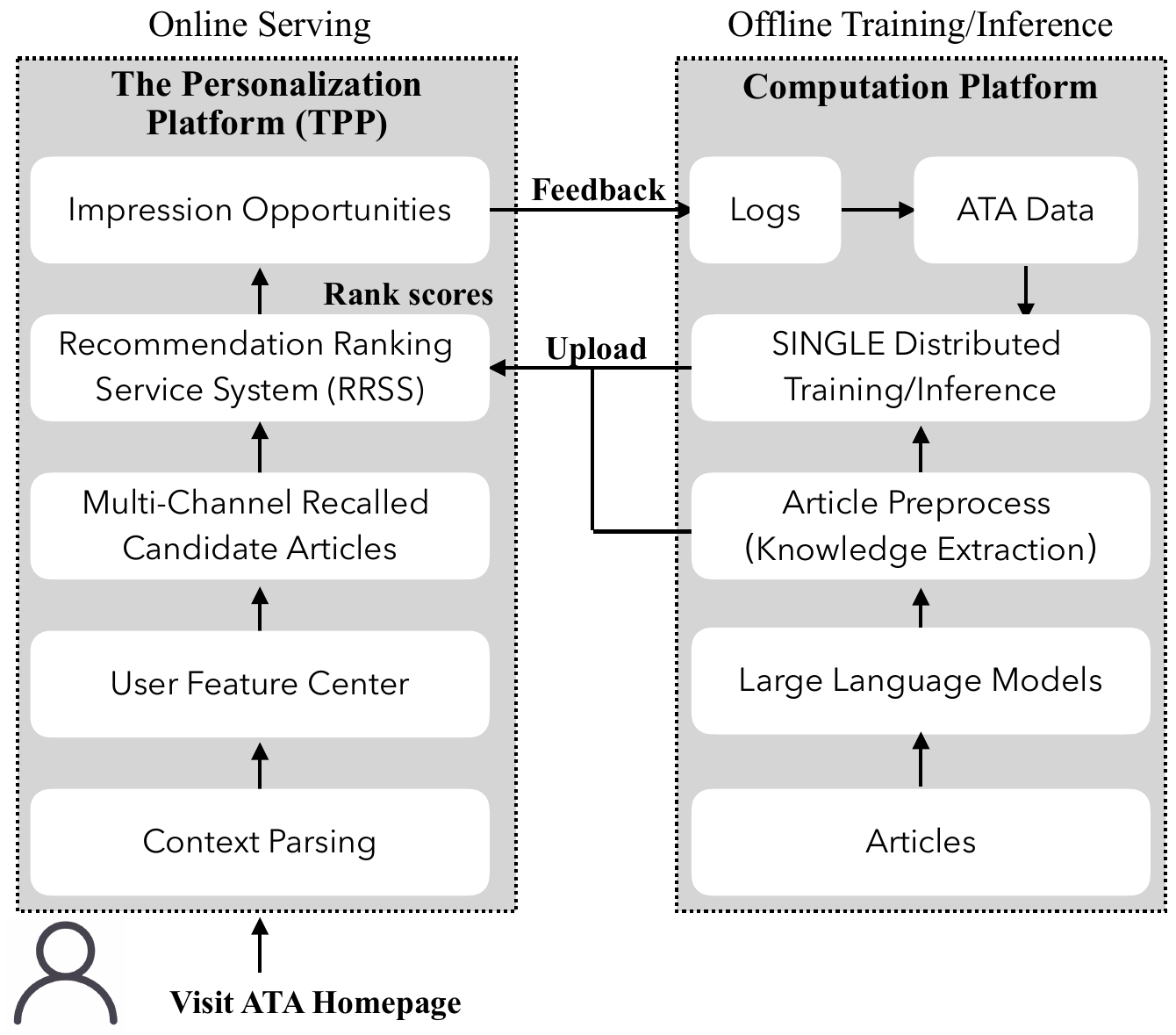}
\caption{The Pipeline of Online System. We deploy SINGLE on the ATA online website of Alibaba.}
\label{fig:system}
\end{figure}
\begin{figure}[t]
\centering
\includegraphics[width=\linewidth]{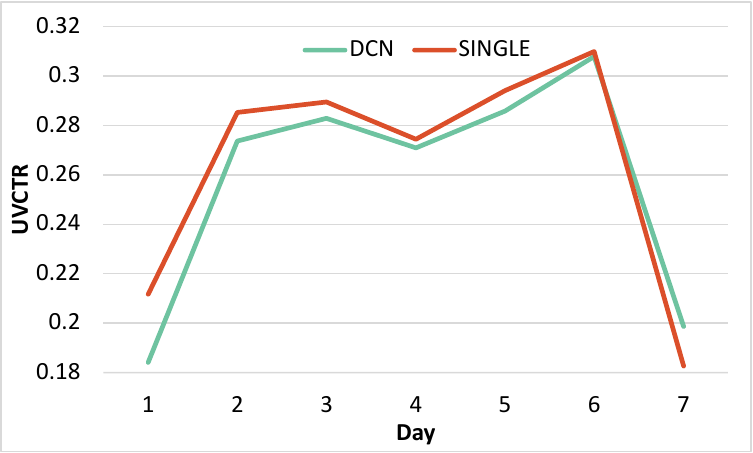}
\caption{Recommendation Effectiveness of Different Methods in the Online A/B Test from Sep. 17, 2023 to Sep. 23, 2023.}
\label{fig:abtest}
\end{figure}
In this part, we further conduct the online A/B test to verify the effectiveness of SINGLE. In detail, we first describe the pipeline of online system development and then show the recommendation performance of DCN~\cite{wang2017deep} and SINGLE. 

As shown in Figure~\ref{fig:system}, the online system is deployed on the ATA platform of Alibaba, which can be divided into the online serving module and offline training/inference module. With the help of the service platform of high-performance computation, the articles can be effectively preprocessed and encoded using SINGLE, which are offline inference processes. Then the user and article representations are uploaded to the online serving system and used for reranking the candidate articles, conducting an efficient recommendation result generation process.

Then we evaluate the effectiveness of online systems in Figure~\ref{fig:abtest}. We choose DCN as our main baseline, which performs better than other baseline models on the ATA dataset (Sec.~\ref{sec:overall}). The online A/B test for article recommendation is conducted at the ATA platform in successive seven days of September 2023. And we evaluate their recommendation performance using the UVCTR metric (Eq.~\ref{eq:uvctr}). To ensure the fairness of comparison, the scheduling engine is modified at the ATA platform to approximately assign $\frac{1}{2}$ daily traffic of personalized ATA homepage to each model during the online A/B test.
Experimental results show that SINGLE achieves on average 2.4$\%$ improvements over the DCN model. This is a significant
improvement in the industrial scenario, which again verifies the
effectiveness of the SINGLE model in recommending the technique articles for users on the ATA platform.

\begin{table*}[t]
\caption{Case Studies. We randomly sample two cases from the ATA dataset to show the effectiveness of our SINGLE model. We show the golden articles, the latest ten user-clicked articles and the top ranked articles by SINGLE. The instant attention scores calculated by Eq.~\ref{eq:attention} are also presented. The matched tokens of user clicked documents and golden articles are \textbf{\emph{\textcolor{red}{emphasized}}}.\label{tab:case}}
\resizebox{\textwidth}{!}{
\begin{tabular}{l|r}
\hline\hline
\multicolumn{2}{l}{\textcolor{blue}{\textbf{Case 1:}} 
 \textbf{\emph{\textcolor{red}{Diffusion Model}}} Interpretation Series Three: \textbf{\emph{\textcolor{red}{Stable-Diffusion}}} \textbf{(Golden Article, R\#2)}} \\\hline
Reading Notes on `The 7 Habits of Highly Effective People'& \cellcolor{red!1}1.0\\
AIGC: Practical Applications of \textbf{\emph{\textcolor{red}{Stable Diffusion}}}& \cellcolor{red!5.3}5.3\\
Exploring Large Models - Making ChatGPT Answer Business & \cellcolor{red!0.3}0.3\\
AIGC Practical Implementation - The Mobile Version of \textbf{\emph{\textcolor{red}{Stable Diffusion}}} That I Developed& \cellcolor{red!0.7}0.7\\
\textbf{\emph{\textcolor{red}{AIGC Visual}}} Introduction and Group Discussion (ChatGPT and \textbf{\emph{\textcolor{red}{Stable Diffusion}}})& \cellcolor{red!2.1}2.1\\
From GAN to \textbf{\emph{\textcolor{red}{Diffusion in Image Generation}}}& \cellcolor{red!0.8}0.8\\
Science Announces the Top Ten Scientific Breakthroughs of the Year& \cellcolor{red!0.3}0.3\\
ReprBERT 2.0: Optimization of Taobao Search Representation BERT Relevance Model& \cellcolor{red!20.2}20.2\\
Endless Imagination: From ChatGPT to AIGC, Research and Reflection on the Implementation of Generative AI& \cellcolor{red!0.7}0.7\\
Collection of the Most Worth-Reading AIGC and ChatGPT Articles for FY23& \cellcolor{red!0.6}0.6\\
Playing with Uncle Da on Prompt Engineering& \cellcolor{red!29.9}29.9\\
\hline
\multicolumn{2}{l}{\textcolor{purple}{\textbf{Top-Ranked Articles:}} 
 Writing an Innovative Proposal Using ChatGPT \textbf{(R\#1)}; LangChain Agent Implementation Principles \textbf{(R\#3)};} \\
 \multicolumn{2}{l}{Useful Prompt Techniques for ChatGPT \textbf{(R\#4)}; A Guide to Building a Budget-Friendly Version of ChatGPT \textbf{(R\#5)}}\\
 \hline\hline

\multicolumn{2}{l}{\textcolor{blue}{\textbf{Case 2:}} Phased Reflection on \textbf{\emph{\textcolor{red}{Medical Search}}} - Retrospective Article \textbf{(Golden Article, R\#1)}} \\\hline
Completing all work using ODPS - Engineering practice of the Quark Risk Control Team & \cellcolor{red!5.2}5.2\\
Have you ever thought about the drawbacks of learning? & \cellcolor{red!2.8}2.8\\
Low-quality title recognition and rewriting & \cellcolor{red!3.8}3.8\\
Application of \textbf{\emph{\textcolor{red}{deep vector retrieval}}} in Fliggy's first-guess content recommendation & \cellcolor{red!9.1}9.1\\
The technology behind duplicate removal of similar web pages - URL normalization algorithm & \cellcolor{red!1.7}1.7\\
Application of fine-grained semantic models in \textbf{\emph{\textcolor{red}{search retrieval}}} & \cellcolor{red!10.5}10.5\\
Intelligent construction practice combining BI and AI, looking at future opportunities for B2B products & \cellcolor{red!2.6}2.6\\
Offline evaluation system for \textbf{\emph{\textcolor{red}{search vector retrieval}}} & \cellcolor{red!7.5}7.5\\
Exploring the capabilities of the LLM model in niche domains: based on Langchain, ChatGLM-6B, and domain knowledge base & \cellcolor{red!4.8}4.8\\
Application of \textbf{\emph{\textcolor{red}{deep vector retrieval}}} in Fliggy's first-guess content recommendation & \cellcolor{red!13.4}13.4\\\hline
\multicolumn{2}{l}{\textcolor{purple}{\textbf{Top-Ranked Articles:}} 
 The Unique Advantages of Reranking in Complex Information Flow Scenarios \textbf{(R\#2)};}\\
\multicolumn{2}{l}{The Evolutionary Journey from GPT to ChatGPT -Beginner's Basic Version \textbf{(R\#3)};} \\
 \multicolumn{2}{l}{Essential Thinking Skills for Programmers: Structured Thinking LangChain Agent Implementation Principles \textbf{(R\#4)}}\\
\hline\hline
\end{tabular}}

\end{table*}
\subsection{Case Studies}
The final experiment conducts some case studies in Table~\ref{tab:case}. We choose two cases from the ATA dataset to show the recommendation effectiveness of SINGLE. Overall, the user instant interest modeling prefers to learn the user characteristics from more recently interacted articles. 

In the first case, the user visits some articles that are about ``Diffusion Model'', and then changes the interest to some technologies of LLMs, such as ``Prompt'' and ``ChatGPT''. Our SINGLE model captures such a topic shift by assigning more attention weights to the previously clicked articles that are about ``BERT Relevance Model'' and ``Prompt Engineering'', which are distinct from previously clicked articles. Thanks to our user constant interest modeling, the recommendation results also cover the article that is about diffusion, showing the effectiveness of user constant interest modeling. The second case shows a different article viewing behavior, which usually keeps the consistent user interest. The user instant interest modeling method also pays more attention to the articles that are about ``retrieval'', showing its effectiveness in filtering out the noise from previously clicked articles.

\section{Conclusion}
In this paper, we propose a u\textbf{S}er view\textbf{ING} f\textbf{L}ow mod\textbf{E}ling (SINGLE) method to mimic the user instant and constant viewing flows and match articles with the user interests. The experimental results show that SINGLE outperforms all previous article recommendation models on MIND and ATA datasets and also achieves significant improvements on the online ATA website of Alibaba. SINGLE thrives on the emergent ability of LLMs to better model the user interests and behaviors for recommendation.

\section*{Acknowledgments}
This work is supported by the Natural Science Foundation of China under Grant No. 62206042 and No. U23B2019, the National Social Science Foundation of China under Grant No. 62137001, and the Joint Funds of Natural Science Foundation of Liaoning Province (No. 2023-MSBA-081).


\bibliographystyle{ACM-Reference-Format}
\balance
\bibliography{sample-base}

\end{document}